\documentclass{mem}
\usepackage{natbib}
\usepackage{graphicx}
\usepackage[a4paper,breaklinks]{hyperref}
\usepackage[export]{adjustbox}
\begin{document}

\title{
Astrochemistry: From primordial gas to present-day clouds
}

\subtitle{}

\author{
Dominik R.G. Schleicher
\inst{1} 
\and 
Stefano Bovino
\inst{2}
\and
Bastian K\"ortgen
\inst{2}
\and
Tommaso Grassi
\inst{3}
\and
Robi Banerjee
\inst{2}
          }

\institute{
Departamento de Astronom\'ia, Universidad de Concepci\'on, Barrio Universitario, Concepci\'on, Chile \email{dschleicher@astro-udec.cl}
\and
Hamburger Sternwarte, Universit\"at Hamburg, Gojenbergsweg 112, 21029 Hamburg, Germany
\and
Centre for Star and Planet Formation, Niels Bohr Institute \& Natural History Museum of Denmark, University of Copenhagen, Oster Voldgade 5-7, DK-1350 Copenhagen, Denmark
}

\authorrunning{Schleicher, Bovino, K\"ortgen \& Grassi}

\titlerunning{Astrochemistry and star formation}

\abstract{Astrochemistry plays a central role during the process of star formation, both in the primordial regime as well as in the present-day Universe. We revisit here the chemistry in both regimes, focusing first on the chemistry under close to primordial conditions, as observed in the so-called Caffau star SDSS J102915+172927, and subsequently discuss deuteration processes in present-day star-forming cores. In models of the high-redshift Universe, the chemistry is particularly relevant to determine the cooling, while it also serves as an important diagnostic in the case of present-day star formation.
\keywords{Astrochemistry -- Stars: formation -- primordial Universe -- molecular clouds}
}
\maketitle{}

\section{Introduction}

Chemistry plays a central role in astrophysics, and particularly during star formation. In the primordial Universe consisting only of hydrogen and helium, the H$_2$ molecule provides the only and relatively inefficient coolant, implying relatively high temperatures of the star-forming clouds compared to Milky Way type conditions. In the absence of dust grains, H$_2$ abundances of order $10^{-3}$  start forming at gas number densities of $10^4$~cm$^{-3}$ due to gas phase reactions \citep{Saslaw67}. The gas becomes fully molecular from densities of about $10^{12}$~cm$^{-3}$ due to three-body reactions \citep{Palla83}. The resulting temperatures of $300$~K or higher are commonly expected to give rise to significantly enhanced masses of the first stars.

The presence of even tiny amounts of dust grains may alter this picture, giving rise to potentially strong fragmentation at high densities \citep{Schneider03}. The star SDSS J102915+172927 is considered as a candidate for such behavior \citep{Klessen12, Schneider12}, as its metal abundances are so low that only dust grains could have contributed relevantly to the cooling \citep{Caffau11}. In the following, we will present a numerical simulation exploring how the formation of such a star may have occurred. We will subsequently turn to deuteration processes in present-day clouds, as explored by \citet{Walmsley04}. We will show that large deuteration fractions can be reached within about a free-fall time. The chemical modeling pursued in this work is based on the publicly available astrochemistry package KROME\footnote{KROME: http://www.kromepackage.org/} \citep{Grassi14}.
 
 \section{The formation of the Caffau star SDSS J102915+172927}
 We model the formation of the Caffau star using the cosmological hydrodynamics code Enzo \citep{Bryan14} combined with the astrochemistry package KROME. We explore the results for two minihalos, one with $10^6$~M$_\odot$ and one with $7\times10^5$~M$_\odot$, forming at $z=22$ and $z=18$, respectively. The halos are part of a cosmological box with size of $300$~kpc~h$^{-1}$, an initial topgrid resolution of $128^3$, two additional nested grids around the halo of choice and 29 levels of refinement. The Jeans length is always resolved with at least $64$ cells. The details of the setup, as well as our treatment of the chemistry and the dust grains, is reported by \citet{Bovino16}. Fixing the metal abundances to the values of the Caffau star, we find that the cooling is predominantly regulated by the depletion factor $f_{\rm dep}=D/Z$, where $D$ denotes the dust-to-gas mass ratio and $Z$ the metallicity.
 
 \begin{center}
\begin{figure*}
\includegraphics[scale=0.4, valign=t]{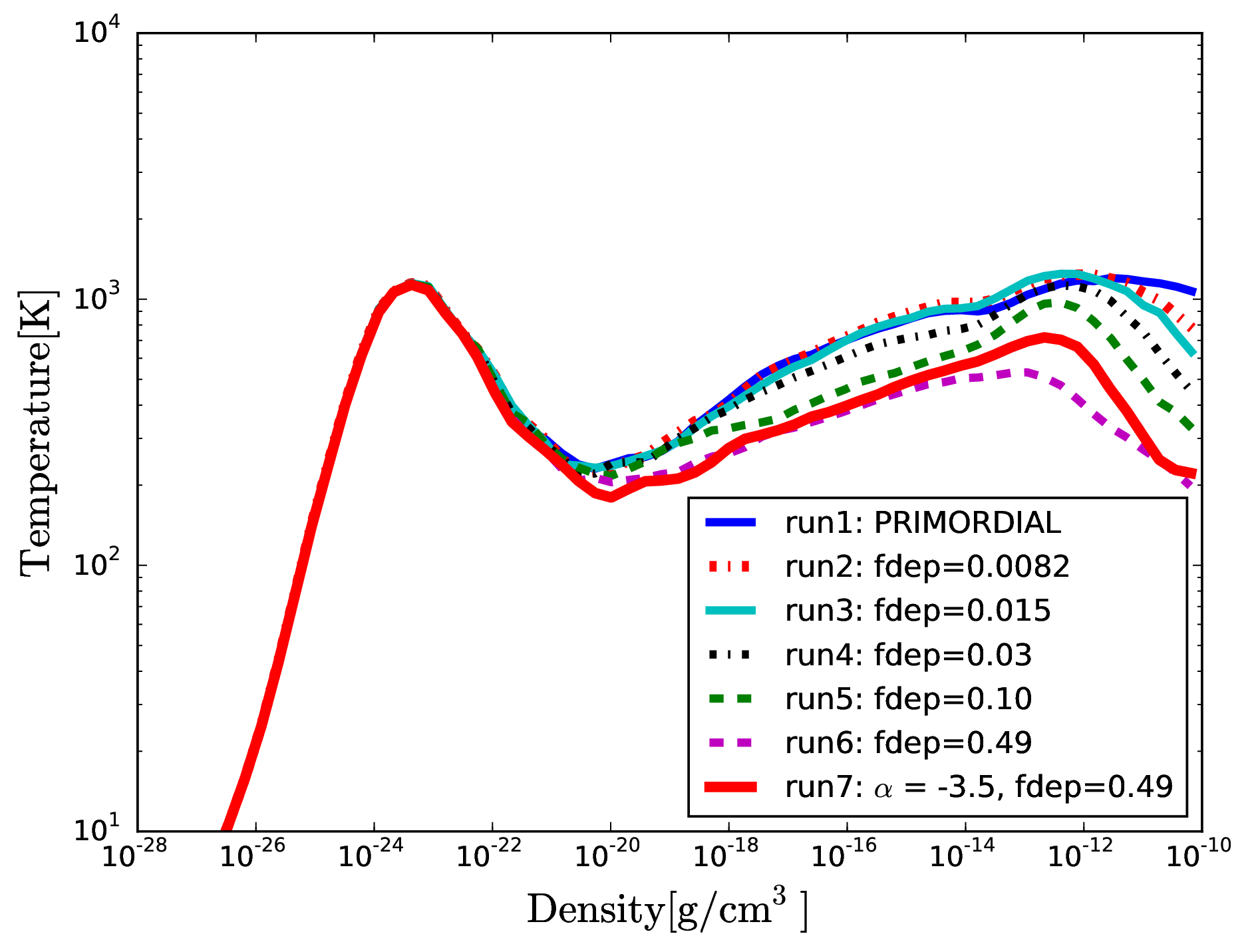}
\includegraphics[scale=0.5, valign=t]{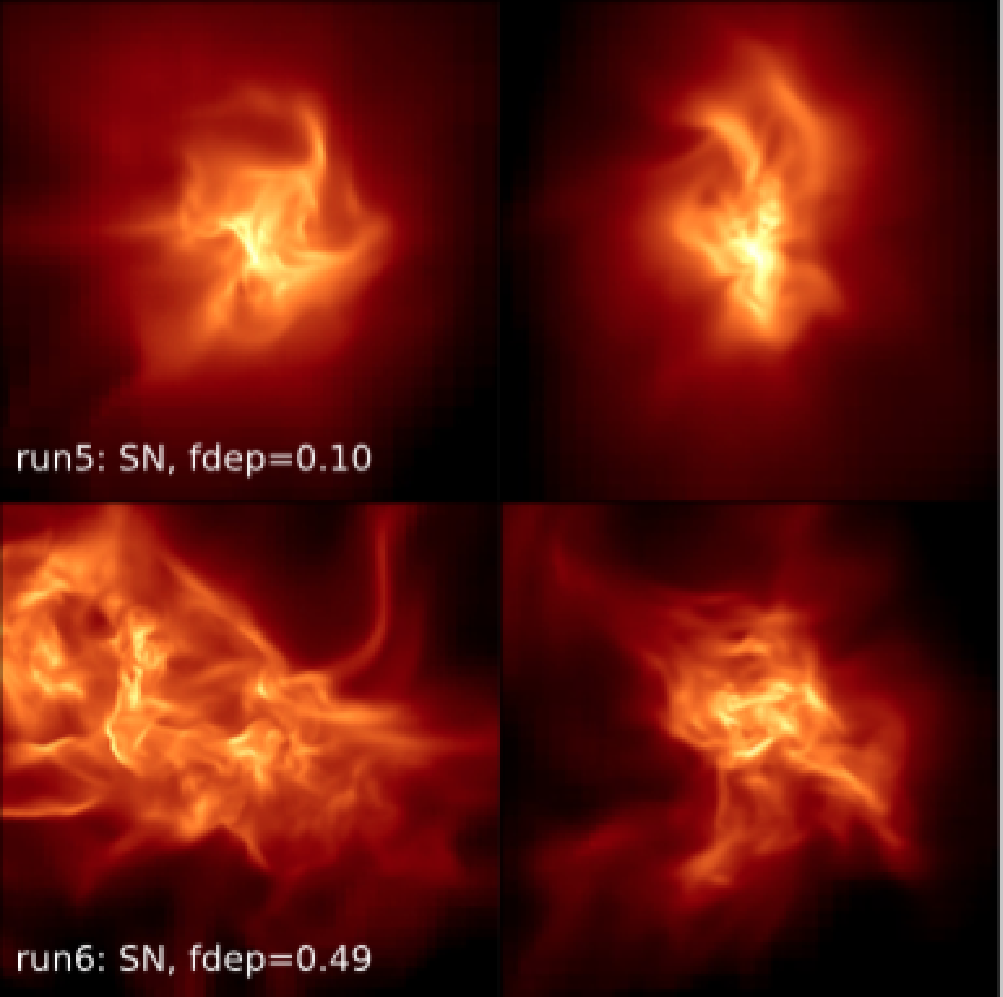}
\caption{Profile of the mass-weighted average temperatures for different dust depletion factors $f_{\rm dep}$ in the dark matter halo with $10^6$~M$_\odot$ (left), as well as the density structure on a scale of 20~AU for selected runs with both halos (left: $10^6$~M$_\odot$, right: $7\times10^5$~M$_\odot$) \citep{Bovino16}. The calculations assume the metal abundances of the Caffau star.}\label{fig:dust}
\end{figure*}
\end{center}
 
 The resulting thermal evolution and the density structure is shown in Fig.~\ref{fig:dust}, where higher depletion factors correspond to lower gas temperatures at high densities. Looking also at the density structure, we found that a strong transition from an approximately spherical collapse mode to a filamentary collapse occurs between depletion factors of $0.1$ and $0.49$, suggesting that the latter corresponds to the critical threshold for fragmentation induced via dust cooling for metallicities as in the Caffau star \citep{Bovino16}.
 
  \section{Deuteration processes during present-day star formation}
  We further explore deuteration processes in prestellar cores, with the aim of inferring the approximate timescale to reach high deuteration fraction. For this purpose, we consider the collapse of a turbulent magnetized Bonnor-Ebert sphere \citep{Ebert55, Bonnor56}, which is modeled with the magneto-hydrodynamical code FLASH \citep{Fryxell00}. The chemistry is modeled using the network of \citet{Walmsley04} under the assumption of full depletion  \citep[see][for a detailed description of the overall setup]{Kortgen17}.
 
\begin{figure*}
        \begin{center}
                \begin{tabular}{ccc}
                        \includegraphics[scale=0.26,angle=-90]{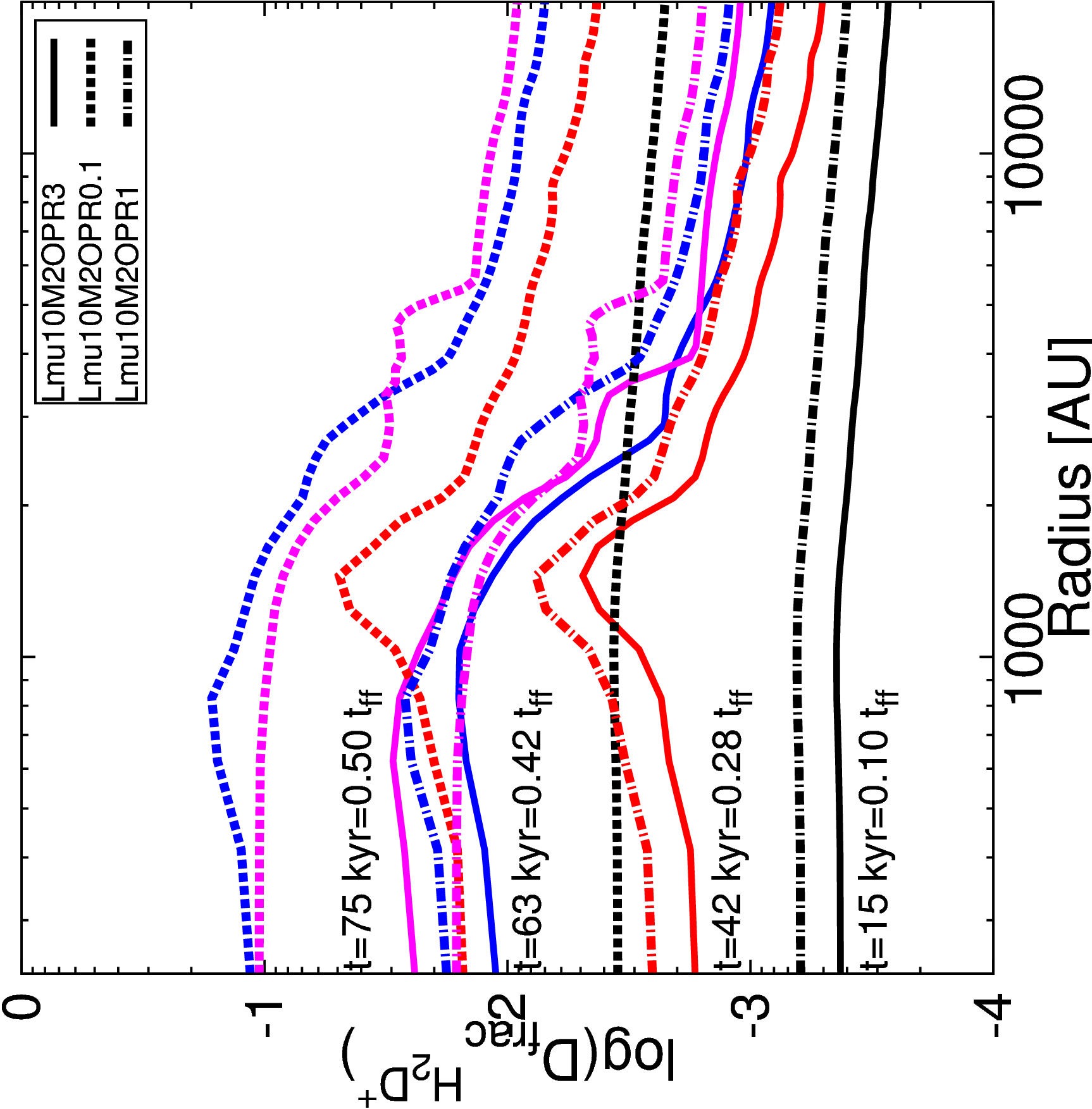}&\includegraphics[scale=0.26,angle=-90]{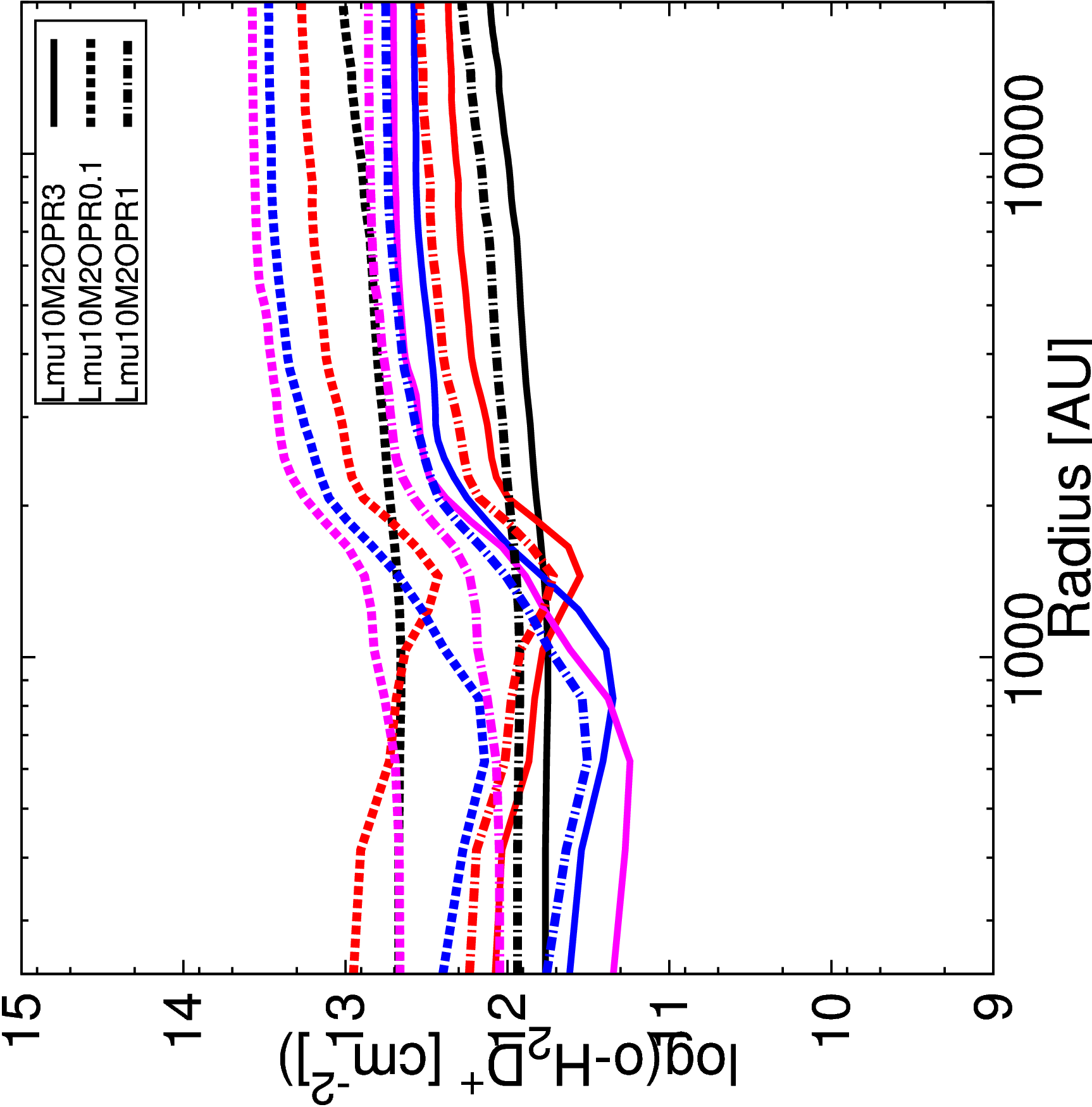}&\includegraphics[scale=0.26,angle=-90]{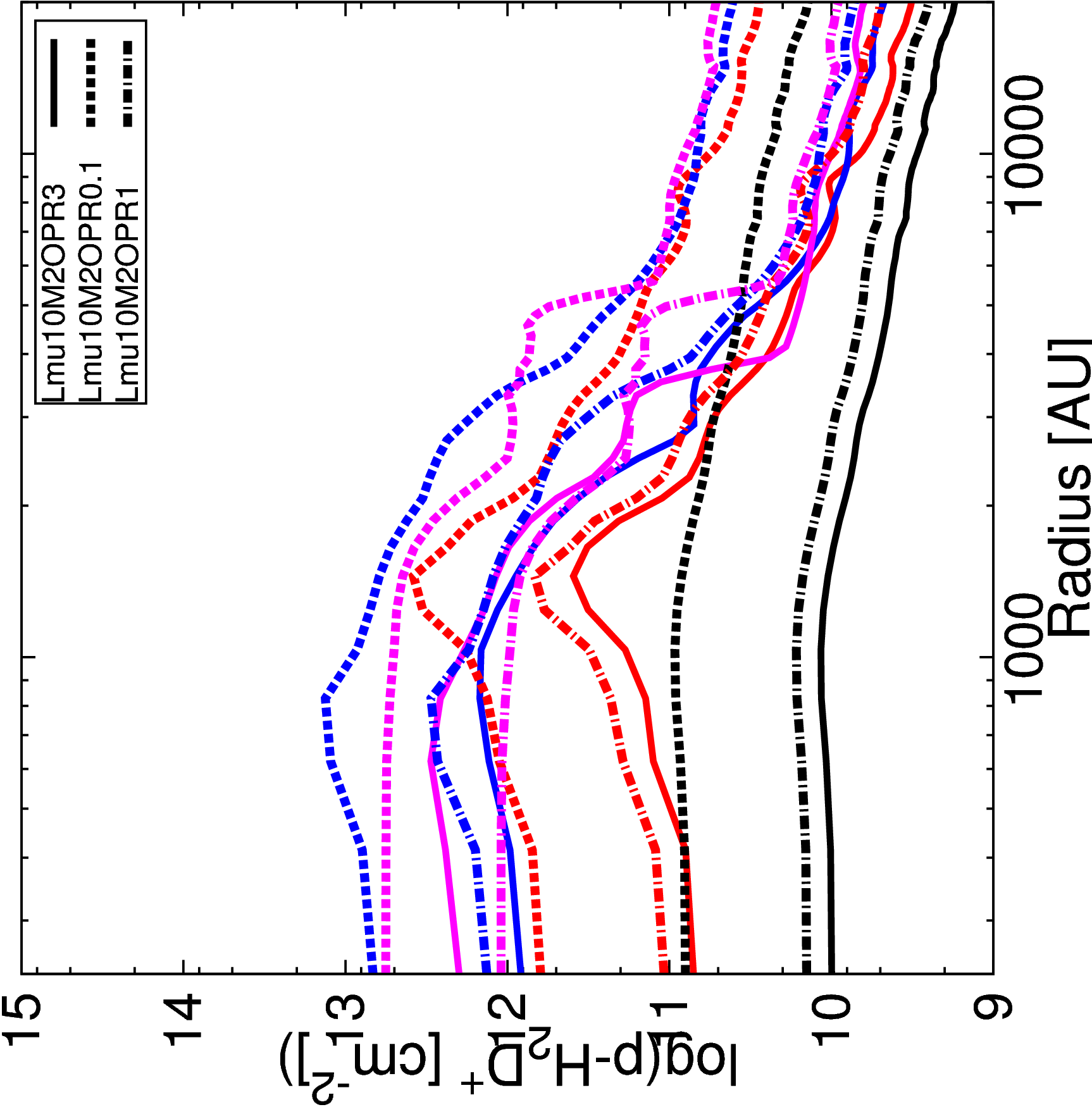}\\
                \end{tabular}
        \end{center}
        \caption{Comparison of radial profiles of the deuteration fraction and the spin states of H$_2$D$^+$  for runs with initial H$_2$ ortho-to-para (OPR) ratio OPR=3 (solid), OPR=1 (dash--dotted) and OPR=0.1 (dashed). See \citet{Kortgen17} for a more detailed discussion.}\label{deuteration}
\end{figure*}

 In Fig.~\ref{deuteration}, we show the radial profiles of the deuteration fraction and the spin states of H$_2$D$^+$ for runs with different initial H$_2$ ortho-to-para ratios, exploring values of 3, 1 and 0.1. Results are given at 15~kyrs, 42~kyrs, 63~kyrs and 75~kyrs, corresponding to $0.1$, $0.28$, $0.42$ and $0.5$ free-fall times. Even in the most unfavorable case with an initial H$_2$ ortho-to-para ratio of $3$, a deuteration fraction of 0.01 is achieved within the central $1000$~AU during half a free-fall time. This demonstrates the  high efficiency of deuteration processes, which we will explore  in more detail in future investigations including finite amounts of depletion.
  
\begin{acknowledgements}
We thank Francesco and Malcolm for the time they spent with us. 
\end{acknowledgements}

%\bibliography{astro}

\begin{thebibliography}{14}
\expandafter\ifx\csname natexlab\endcsname\relax\def\natexlab#1{#1}\fi

\bibitem[{{Bonnor}(1956)}]{Bonnor56}
{Bonnor}, W.~B. 1956, \mnras, 116, 351

\bibitem[{{Bovino} {et~al.}(2016){Bovino}, {Grassi}, {Schleicher}, \&
  {Banerjee}}]{Bovino16}
{Bovino}, S., {Grassi}, T., {Schleicher}, D.~R.~G., \& {Banerjee}, R. 2016,
  \apj, 832, 154

\bibitem[{{Bryan} {et~al.}(2014){Bryan}, {Norman}, {O'Shea}, {Abel}, {Wise},
  {Turk}, {Reynolds}, {Collins}, {Wang}, {Skillman}, {Smith}, {Harkness},
  {Bordner}, {Kim}, {Kuhlen}, {Xu}, {Goldbaum}, {Hummels}, {Kritsuk}, {Tasker},
  {Skory}, {Simpson}, {Hahn}, {Oishi}, {So}, {Zhao}, {Cen}, {Li}, \& {Enzo
  Collaboration}}]{Bryan14}
{Bryan}, G.~L., {Norman}, M.~L., {O'Shea}, B.~W., {et~al.} 2014, \apjs, 211, 19

\bibitem[{{Caffau} {et~al.}(2011){Caffau}, {Bonifacio}, {Fran{\c c}ois},
  {Sbordone}, {Monaco}, {Spite}, {Spite}, {Ludwig}, {Cayrel}, {Zaggia},
  {Hammer}, {Randich}, {Molaro}, \& {Hill}}]{Caffau11}
{Caffau}, E., {Bonifacio}, P., {Fran{\c c}ois}, P., {et~al.} 2011, \nat, 477,
  67

\bibitem[{{Ebert}(1955)}]{Ebert55}
{Ebert}, R. 1955, \zap, 37, 217

\bibitem[{{Fryxell} {et~al.}(2000){Fryxell}, {Olson}, {Ricker}, {Timmes},
  {Zingale}, {Lamb}, {MacNeice}, {Rosner}, {Truran}, \& {Tufo}}]{Fryxell00}
{Fryxell}, B., {Olson}, K., {Ricker}, P., {et~al.} 2000, \apjs, 131, 273

\bibitem[{{Grassi} {et~al.}(2014){Grassi}, {Bovino}, {Schleicher}, {Prieto},
  {Seifried}, {Simoncini}, \& {Gianturco}}]{Grassi14}
{Grassi}, T., {Bovino}, S., {Schleicher}, D.~R.~G., {et~al.} 2014, \mnras, 439,
  2386

\bibitem[{{Klessen} {et~al.}(2012){Klessen}, {Glover}, \& {Clark}}]{Klessen12}
{Klessen}, R.~S., {Glover}, S.~C.~O., \& {Clark}, P.~C. 2012, \mnras, 421, 3217

\bibitem[{{K{\"o}rtgen} {et~al.}(2017){K{\"o}rtgen}, {Bovino}, {Schleicher},
  {Giannetti}, \& {Banerjee}}]{Kortgen17}
{K{\"o}rtgen}, B., {Bovino}, S., {Schleicher}, D.~R.~G., {Giannetti}, A., \&
  {Banerjee}, R. 2017, \mnras, 469, 2602

\bibitem[{{Palla} {et~al.}(1983){Palla}, {Salpeter}, \& {Stahler}}]{Palla83}
{Palla}, F., {Salpeter}, E.~E., \& {Stahler}, S.~W. 1983, \apj, 271, 632

\bibitem[{{Saslaw} \& {Zipoy}(1967)}]{Saslaw67}
{Saslaw}, W.~C. \& {Zipoy}, D. 1967, \nat, 216, 976

\bibitem[{{Schneider} {et~al.}(2003){Schneider}, {Ferrara}, {Salvaterra},
  {Omukai}, \& {Bromm}}]{Schneider03}
{Schneider}, R., {Ferrara}, A., {Salvaterra}, R., {Omukai}, K., \& {Bromm}, V.
  2003, \nat, 422, 869

\bibitem[{{Schneider} {et~al.}(2012){Schneider}, {Omukai}, {Limongi},
  {Ferrara}, {Salvaterra}, {Chieffi}, \& {Bianchi}}]{Schneider12}
{Schneider}, R., {Omukai}, K., {Limongi}, M., {et~al.} 2012, \mnras, 423, L60

\bibitem[{{Walmsley} {et~al.}(2004){Walmsley}, {Flower}, \& {Pineau des
  For{\^e}ts}}]{Walmsley04}
{Walmsley}, C.~M., {Flower}, D.~R., \& {Pineau des For{\^e}ts}, G. 2004, \aap,
  418, 1035

\end{thebibliography}
%\bibliographystyle{aa}

\end{document}